\documentclass{article}

\usepackage{array}
\usepackage{PRIMEarxiv}
\usepackage{subfigure}	
\usepackage[utf8]{inputenc} 
\usepackage[T1]{fontenc}    
\usepackage{hyperref}       
\usepackage{url}            
\usepackage{booktabs}       
\usepackage{amsfonts}       
\usepackage{nicefrac}       
\usepackage{microtype}      
\usepackage{lipsum}
\usepackage{fancyhdr}       
\usepackage{graphicx}       
\usepackage{amsmath}
\graphicspath{{media/}}     

\title{Predict stock prices with ARIMA and LSTM 
}

\author{
  Ruochen Xiao\textsuperscript{1}  \\
  Department of Computer Science, Shanghai University\\
  \texttt{\{Ruochen Xiao\}ninaxiao1208@shu.edu.cn} \\
  \And
  Yingying Feng\textsuperscript{2} \\
  Department of East Asian Business, The university of Sheffield \\
  \AND
  Lei Yan\textsuperscript{3} \\
  Department of Biology, Zhejiang University \\
  \AND
  Yihan Ma\textsuperscript{3} \\
  Department of Mathematics, University of Chinese Academy of Sciences \\
}

\begin{document}

\maketitle

\begin{abstract}
MAE, MSE and RMSE performance indicators are used to analyze the performance of different stocks predicted by LSTM and ARIMA models in this paper. 50 listed company stocks from finance.yahoo.com are selected as the research object in the experiments. The dataset used in this work consists of the highest price on transaction days, corresponding to the period from 01 January 2010 to 31 December 2018. For LSTM model, the data from 01 January 2010 to 31 December 2015 are selected as the training set, the data from 01 January 2016 to 31 December 2017 as the validation set and the data from 01 January 2018 to 31 December 2018 as the test set. In term of ARIMA model, the data from 01 January 2016 to 31 December 2017 are selected as the training set, and the data from 01 January 2018 to 31 December 2018 as the test set. For both models, 60 days of data are used to predict the next day. After analysis, it is suggested that both ARIMA and LSTM models can predict stock prices, and the prediction results are generally consistent with the actual results;and LSTM has better performance in predicting stock prices(especially in expressing stock price changes), while the application of ARIMA is more convenient.

\end{abstract}


\section{Introduction}
In the fields of statistics and probability theory as well as machine learning, various methods have been developed to predict the value of the stock market index. In the last few years, there have been incredible success applying RNNs(Recurrent Neural Network) to a variety of problems: speech recognition, language modeling, translation, image captioning......Essential to these successes is the use of “LSTMs,” a very special kind of recurrent neural network which works, for many tasks, much better than the standard version\cite{en3}.

 Stock price data have the characteristics of time series, and the Auto-Regressive Integrated Moving Average (ARIMA) approach is often used to predict time series. The basic idea of the ARIMA model: the data series of the predicted object over time is considered as a random sequence, and a certain mathematical model is used to describe this sequence approximately. The Auto-Regressive Integrated Moving Average (ARIMA)\cite{en2} approach is used very widely, because it can obtain useful statistical properties. It is also very flexible as it can represent multiple different time series using different order parameters. 
 
At the same time, based on machine learning long short-term memory (LSTM)\cite{en1} which has the advantages of analyzing relationships among time series data through its memory function, we propose a forecasting method of stock price based on LSTM. The LSTM technique is a model that extends RNN (Recurrent Neural Network) memory. In addition, the main difference between LSTM and RNN is that it adds a cell to the algorithm to determine whether the information is useful or not. Typically, repetitive neural networks have "short-term memory" because they use persistent prior knowledge for use in the existing neural network. Essentially, previous information is used in the current task.Then the forecasting results of these models are analyzed and compared. The data utilized in Google Apple Netflix and Amazon.
 
The novelty of the paper lies in analyzing the similarities and differences between the LSTM and ARIMA models and comparing the results and running speed of stock forecasting under these two models, filling a relevant gap in the field.

\section{Classical Time Series (ARIMA)}
\label{sec:headings}
\subsection{Concept}
ARIMA model, the full name of Autoregressive Integrated Moving Average Model (Autoregressive Integrated Moving Average Model), is a time series forecasting method proposed by Box and Jenkins in the early 1970s. ARIMA model refers to the model established by regressing the dependent variable only on its lag value and the present value and lag value of the random error term in the process of converting a non-stationary time series into a stationary time series.
\subsection{Principle}
The basic idea of the ARIMA model is to treat the data sequence formed by the prediction object over time as a random sequence, and use a certain mathematical model to approximately describe the sequence. Once identified, this model can predict future values from past and present values of the time series. Modern statistical methods and econometric models have been able to help companies predict the future to some extent.
\subsection{Math Model}
The ARIMA model has three parameters: p, d, q.

p: lags of the time series data itself used in the forecast model (lags), also called AR/Auto-Regressive item.\newline
d: the time series data needs to be differentiated in several orders  to be stable, also called the Integrated item.\newline
q: the number of lags (lags) of the forecast error used in the forecast model, also known as the MA/Moving Average term.\newline

\begin{equation}
\left(1-\sum_{i=1}^{p} \phi_{i} L^{i}\right)(1-L)^{d} X_{t}=\left(1+\sum_{i=1}^{q} \theta_{i} L^{i}\right) \epsilon_{t}
\end{equation}
$\psi$ is AR parameter, $\theta$ is MA parameter, L is lag operator, d is a positive integer.

\section{Deep Learning Time Series (LSTM)}

\subsection{The Concept of LSTM}

Recurrent neural network (RNN) can transmit historical information through a neural network architecture similar to chain (Zhou et al., 2015). When processing sequential data, it will check the current input i (t) and the output h (t-1) of the previous hidden state in each time period (Zhou et al., 2015). However, when the gap between the two time steps becomes large, the standard RNNs becomes unable to learn long-term dependence (Zhou et al., 2015). Therefore, LSTM cell are proposed by Hochreiter and Schmidhuber (1997) to deal with the problem of "long-term dependence": the memory capacity of standard recurrent cell is improved by introducing a "gate" into the cell. It is worth mentioning that LSTM is specifically designed for time series data (Zhou et al., 2015). 

\subsection{The Original LSTM}

At present, many researchers have modified and popularized LSTM (Gers, 2001; Gers \&Schmidhuber,2000), derived from LSTM with a forget gate, without a forget gate, with a peephole connection. Generally, the term LSTM cell is understood as LSTM with a forget gate, whereas the original LSTM model had only input and output gates (Yu et al., 2019), as shown in Figure.1:

\begin{figure}
  \centering
  \includegraphics[width=0.7\textwidth]{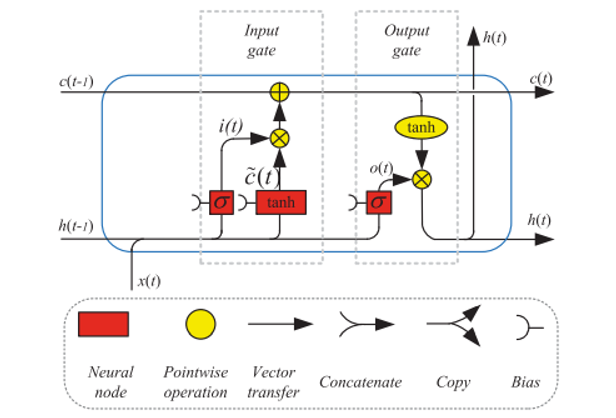} 
  \caption{Original LSTM Architecture (Yu et al., 2019)} 
  \label{Fig.1} 
\end{figure}

Specifically, the mathematical expression of LSTM in Figure. 1 can be written as follows:

\begin{equation}
\begin{array}{l}
f_{t}=\sigma\left(W_{f h} h_{t-1}+W_{f x} x_{t}+b_{f}\right) \\
i_{t}=\sigma\left(W_{i h} h_{t-1}+W_{i x} x_{t}+b_{i}\right) \\
\bar{c}_{t}=\tanh \left(W_{\mathrm{ch}} h_{t-1}+W_{\bar{r} r} x_{t}+b_{\bar{r}}\right) \\
c_{t}=f_{t} \cdot c_{t-1}+i_{t} \cdot \bar{c}_{t} \\
o_{t}=\sigma\left(W_{c h} h_{t-1}+W_{o x} x_{t}+b_{o}\right) \\
h_{t}=o_{t} \cdot \tanh \left(c_{t}\right)
\end{array}
\end{equation}

Where ct represents the cell state of LSTM. Wi, Wc˜, and Wo are the weights and the operator '.' represents the multiplication of the two vectors. When updating the cell state, the input gate determines what data can be stored in the cell state, while the output gate determines what data can be output based on the cell state.

\subsection{The Principle of LSTM}

Gers, Schmidhuber, and Cummins (2000) modified the original LSTM by introducing a forgetting gate into the unit, as shown in Figure 2:
\begin{figure}
\centering
\includegraphics[width=0.7\textwidth]{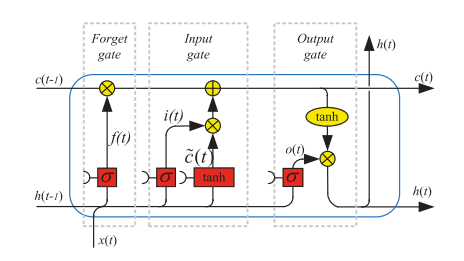} 
\caption{ Architecture of LSTM with a forget gate (Yu et al., 2019)} 
\label{Fig.2} 
\end{figure}

The key principles of an LSTM cell with a forget gate revolve around organizing its internal operations according to two qualitatively different but co-operative objectives: data and data control. The data component is responsible for candidate data signals (range -1 to 1), while the control component is responsible for control signals (range 0 to 1). The candidate data signal is multiplied by the control signal to allocate the partial amount of candidate data that is allowed to propagate to a predetermined node in the cell. Therefore, when the forgetting gate has a value of 1, it retains all data; Meantime, a value of 0 means that it will discard all data. Similarly, for the intermediate value of the control signal (in the range between 0 and 1), the corresponding percentage of the data will be supplied to the next function in the cell.

\subsection{The Mathematical Expression of LSTM}

The mathematical expression of an LSTM cell with a forget gate can be written as follows:

\begin{equation}
\begin{array}{l}
f_{t}=\sigma\left(W_{f h} h_{t-1}+W_{f x} x_{t}+b_{f}\right) \\
i_{t}=\sigma\left(W_{i h} h_{t-1}+W_{i x} x_{t}+b_{i}\right) \\
\tilde{c}_{t}=\tanh \left(W_{\mathrm{ch}} h_{t-1}+W_{\bar{r}} x_{t}+b_{\bar{r}}\right) \\
c_{t}=f_{t} \cdot c_{t-1}+i_{t} \cdot \bar{c}_{t} \\
o_{t}=\sigma\left(W_{c h} h_{t-1}+W_{o x} x_{t}+b_{o}\right) \\
h_{t}=o_{t} \cdot \tanh \left(c_{t}\right)
\end{array}
\end{equation}

$\sigma$ is a logarithmic sigmoid function whose output is at [0, 1], tanh is a hyperbolic tangent function whose output is at [-1, 1], and “·“ represents element multiplication.

Jozefowicz et al. (2015) found that when the bias of the forget gate was increased, the performance of bf and LSTM network would generally become better. In addition, Schmidhuber et al. (2007) proposed that LSTM can sometimes be trained by combining evolutionary algorithms with other techniques, rather than by pure gradient descent.

\section{Experiment}

\subsection{Data splitting}
In our experiments, we selected 50 listed company stocks from finance.yahoo.com as the research object. The dataset used in this work consists of the highest price on transaction days, corresponding to the period from 01 January 2010 to 31 December 2018. In addition, for LSTM model, we select the data from 01 January 2010 to 31 December 2015 as the training set, the data from 01 January 2016 to 31 December 2017 as the validation set and the data from 01 January 2018 to 31 December 2018 as the test set. At the same time, in term of ARIMA model, we select the data from 01 January 2016 to 31 December 2017 as the training set, and the data from 01 January 2018 to 31 December 2018 as the test set. For both models, we used 60 days of data to predict the next day. That is, 60 inputs and 1 output at a time. The following two figures are two examples of dividing data sets for LSTM and ARIMA models using Google as an example.

\begin{figure} 
\centering 
\includegraphics[width=0.7\textwidth]{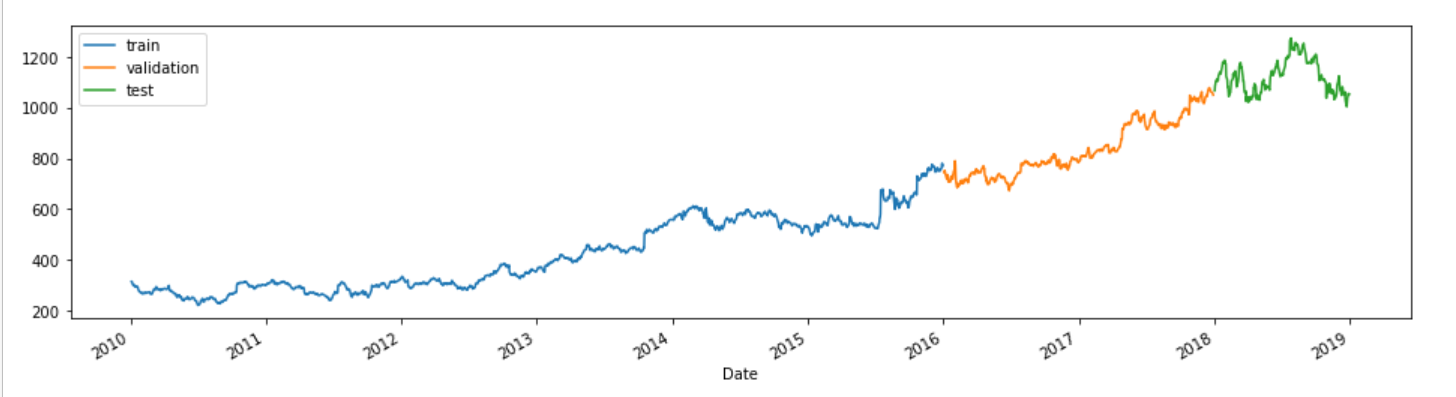} 
\caption{Data splitting on Google for LSTM model} 
\label{Fig.3} 
\end{figure}

\begin{figure} 
\centering 
\includegraphics[width=0.7\textwidth]{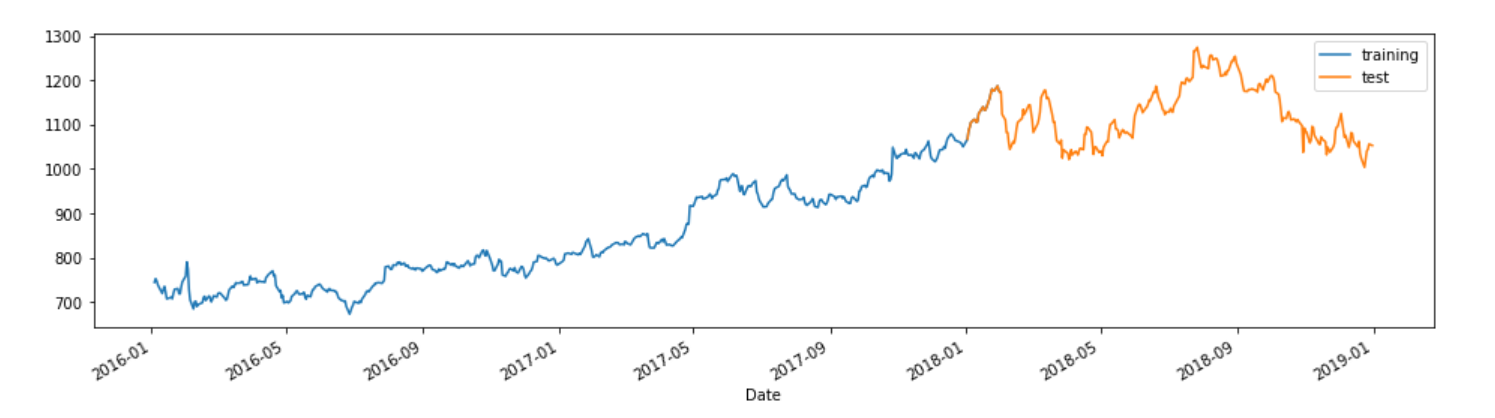} 
\caption{Data splitting on Google for ARIMA model} 
\label{Fig.4} 
\end{figure}

\subsection{ARIMA model}
ARIMA is shorthand for Auto Regressive Integrated Moving Average model. ARIMA is used for the analysis of stationary sequences or sequences that are stationary by difference, and is abbreviated as ARIMA(p, d, q). It's expressed by the formula:
\begin{equation}
\begin{array}{l}
\quad \Delta \mathrm{d} \mathrm{Z}_{\mathrm{t}}=\mathrm{X}_{1}=\varphi_{1} \mathrm{X}_{1-1}+\varphi_{2} \mathrm{X}_{1-2}+\cdots+\varphi_{\mathrm{p}} \mathrm{X}_{1-\mathrm{p}}+\mathrm{a}_{1}-\theta_{1} \mathrm{a}_{1-1}-\theta_{2} \mathrm{a}_{1-2} \\
\cdots \cdots-\theta_{\mathrm{q}} \mathrm{a}_{1-\mathrm{q}}
\end{array}
\end{equation}
Where p, d and q are autoregressive order, difference order and moving average order respectively. $\mathrm{Z}_{\mathrm{t}}$ is time series; $\mathbf{X}_{\mathbf{t}}$ is the time series value after d-order difference; $a_{t-q}$ is the random disturbance term with time t-q; $\varphi_{p} , \theta_{q}$ are the coefficients before the corresponding terms.
\subsubsection{Check for stationarity}

Since ARMA and ARIMA require time series to meet the requirements of stationarity and non-white noise, difference method and smoothing method (rolling average and rolling standard deviation) are used to realize the stationarity operation of sequences. In general, the stationarity of time series can be achieved by first order difference method, and sometimes second order difference is required.
\begin{figure}
	\centering
	\subfigure[first order difference] {\includegraphics[width=.3\textwidth]{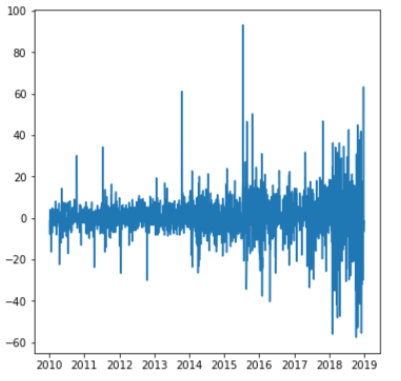}}
	\subfigure[second order difference] {\includegraphics[width=.3\textwidth]{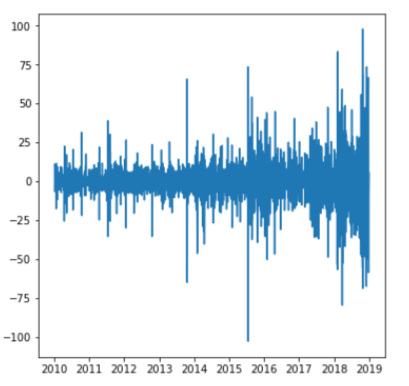}}
	\caption{Difference analysis}
	\label{fig_E1}
\end{figure}

\subsubsubsection{ADF test}
The above two figures are the first-order difference and second-order difference of the highest price of stock in the training set data respectively. It can be seen from the figure that the first-order difference basically meets the need of stationarity, and so does the second-order difference. After the first difference of the sequence, ADF test is performed. The Test results show that the Value of Test Statistic is -15.215325245, which is less than the Critical Value given by Critical Value at the significance level of 1$\%$,5$\%$ and 10$\%$, and p-value= 2.453635256346e-31<0.05; Therefore, the null hypothesis is rejected, indicating that the highest order sequence is stable after first-order difference.

\subsubsection{Parameter estimation}

ARIMA model contains three parameters: p, d and q.

p: Indicates the Regressive number of time series data used in the prediction model, also known as AR/ auto-regressive.

d: Indicates that the time series data needs several order differentiation to be stable, which is also called Integrated term.

q: represents the lag number of prediction error adopted in the prediction model, also known as MA/Moving Average term.
\subsubsubsection{Confirmation of parameter d}
If a time series is stationary after difference operation, it is a differential stationary series and can be analyzed by using ARIMA model. In other words, if a stationary sequence is obtained after d difference is made on the time series, the ARIMA(p,d,q) model can be used, where d is the number of difference.

\hspace*{\fill} \\
First, through THE ADF test, check the stationarity of the original time series, if the original time series is stationary, then d=0; If the original data is not stable, then the difference calculation is carried out and the ADF test is passed until the time series is stable. Generally, the number of difference does not exceed 2 times.

\hspace*{\fill} \\
After first-order difference, the data in this paper is stable through ADF test, so we set d as 1.

\subsubsubsection{Confirmation of parameter p and q}
Generally, in time series analysis, autocorrelation function (ACF) and partial autocorrelation function (PACF) are used to distinguish coefficients and orders of ARMA(P, Q) model. However, the grading method of model by tailing and truncating is often very subjective. In the selection of ARMA parameters, the AIC criterion and BIC criterion can effectively compensate for the subjectivity of order determination based on autocorrelation graph and partial autocorrelation graph, and help us find the relative optimal fitting model within a limited range of order.

\hspace*{\fill} \\

1.AIC

AIC Criterion was proposed by Japanese statistician Akaike and 1973, and its full name is Akaike Information Criterion. It is a weighting function of the fitting accuracy and the number of parameters:

\hspace*{\fill} \\
AIC=2 (number of model parameters) -2ln (maximum likelihood function of model) 

\hspace*{\fill} \\

2.BIC

AIC provides effective rules for model selection, but it also has some shortcomings. When the sample size is large, the fitting error of the information provided in the AIC criterion will be amplified by the sample size, the number of parameters of penalty factor but has nothing to do with sample size (2) has been, so when the sample size is large, using AIC criterion to select the model and real convergence, it is usually better than the real model of the number of unknown parameters. BIC (Bayesian Information Criterion) is a discriminant criterion proposed by Schwartz in 1978 according to Bayes theory, called BIC criterion, which makes up for AIC. BIC is defined as:

\hspace*{\fill} \\
BIC = ln(n)(number of parameters in the model) -2ln (maximum likelihood function value of the model)

\hspace*{\fill} \\
In this paper, the best combination of P and Q in our model is searched by a grid search method. Here, BIC is used for the test, and the thermal map is drawn as follows:
\begin{figure} 
\centering 
\includegraphics[width=0.7\textwidth]{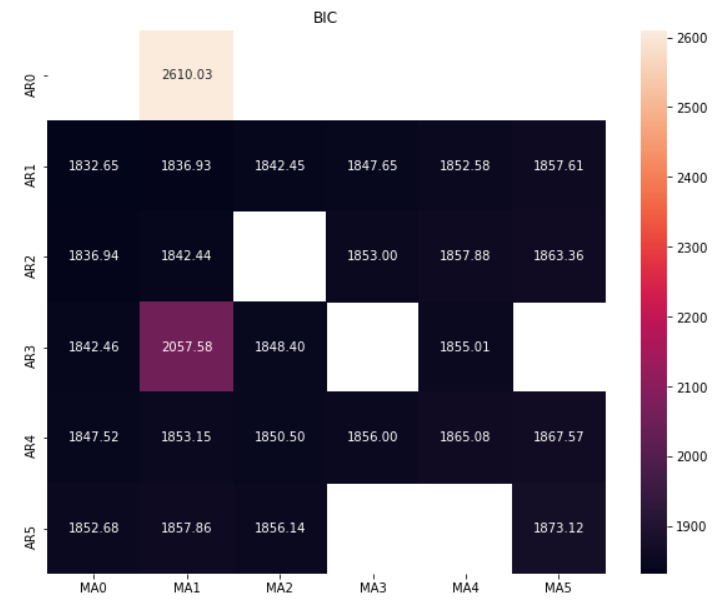} 
\caption{BIC thermal map} 
\label{Fig.6} 
\end{figure}

In fact, we can get the optimal values of p and q in a simpler way, which is sm.tsa.armaOrderSelectic in SM. The result is as follows: AIC (1, 0) BIC (1, 0) indicates that AR(1) model should be selected. Generally speaking, the order of ARMA model obtained by BIC criterion is lower than that of AIC.

\subsubsubsection{Model test}
There are two main model tests here: 1) to test the significance of parameter estimation (T test) 2) to test the randomness of residual sequence, that is, the residuals are independent.

The randomness of the residual sequence can be tested by the autocorrelation function method, that is, the autocorrelation function of the residual.
\begin{figure} 
\centering 
\includegraphics[width=0.7\textwidth]{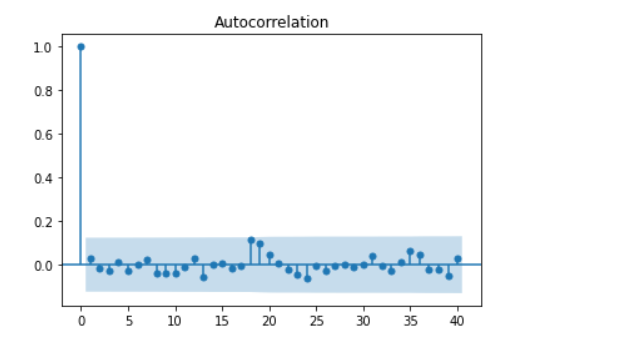} 
\caption{autocorrelation} 
\label{Fig.7} 
\end{figure}
\hspace*{\fill} \\
According to the above conclusions, the final model established in this paper is ARIMA(1,0,1). Here, the establishment of ARIMA model is finished. The next section will introduce the prediction results of ARIMA.

\subsection{LSTM model}

\subsubsection{Normalization}
To detect stock price pattern, it is necessary to normalize the stock price data. Since the LSTM neural network requires the stock patterns during training, we use “min-max” normalization method to reform dataset, which keeps the pattern of the data, as follow:

\begin{equation}
X_{t}^{n}=\frac{X_{t}-\min \left(X_{t}\right)}{\max \left(X_{t}\right)-\min \left(X_{t}\right)}
\end{equation}
where $X_{t}^{n}$ denotes the data after normalization. Accordingly, de-normalization is required at the end of the prediction process to get the original price, which is given by
\begin{equation}
\hat{X}_{t}=\hat{X}_{t}^{n}\left[\max \left(X_{t}\right)-\min \left(X_{t}\right)\right]+\min \left(X_{t}\right)
\end{equation}
where $\hat{X}_{t}$ denotes the predicted data and $\hat{X}_{t}$ denotes the predicted data after de-normalization.
Note that compound score is not normalized, since the compound score range from -1 to 1, which means all the compound score data has the same scale, so it is not require the normalization processing.

\subsubsection{Training Setting}

Parameter adjustment is the key to train neural network. In this paper, the optimal parameters are obtained by comparing different parameters to obtain the deviation of the predicted results. The adjusted parameters are respectively Dropout, neural network layer number, and neural network node number.

\begin{table}[hbp]
 \centering
 \caption{Deviation results of different Dropout}
 \label{tab:pagenum}
 \begin{tabular}{llll}
  \toprule
               & Train Score & Validation Score & Test Score \\
  \midrule
  Dropout = 0.1& 0.0002576723927631974 & 0.0006482924218289554 & 0.00517316022887826\\
  Dropout = 0.2& 0.0006563749047927558 & 0.0011541662970557809 & 0.006828213110566139\\
  Dropout = 0.5& 0.0004621293337550014 & 0.0013630003668367863& 0.008681918494403362\\
  
  \bottomrule
 \end{tabular}
\end{table}
\begin{table}[hbp]
 \centering
 \caption{The deviation results of different LSTM layers}
 \label{tab:pagenum}
 \begin{tabular}{llll}
  \toprule
               & Train Score & Validation Score & Test Score \\
  \midrule
  Layers = 3& 0.0002576723927631974 & 0.0006482924218289554 & 0.00517316022887826\\
  Layers = 4& 0.000796486740000546 & 0.0015846733003854752 & 0.006062280386686325\\

  \bottomrule
 \end{tabular}
\end{table}

\begin{table}[hbp]
 \centering
 \caption{The deviation results of different LSTM layers}
 \label{tab:pagenum}
 \begin{tabular}{llll}
  \toprule
               & Train Score & Validation Score & Test Score \\
  \midrule
  Units = 100& 0.0002576723927631974 & 0.0006482924218289554 & 0.00517316022887826\\
  Units = 200& 0.0003175494493916631 & 0.0006133969291113317 & 0.0049438211135566235\\

  \bottomrule
 \end{tabular}
\end{table}

In this paper, the LSTM model has seven layers, followed by an LSTM layer, a dropout layer, an LSTM layer, a dropout layer, an LSTM layer, a dropout layer, a dense layer, respectively. The dropout layers (with dropout rate 0.1) prevent the network from overfitting. The unit number for each LSTM layer is set 100. The dense layer is used to reshape the output. Since a network will be difficult to train if it contains a large number of LSTM layers, we use three LSTM layers here. 
In each LSTM layer, the loss function is the cross entropy error:
\begin{equation}
E=-\sum_{k} t_{k} \log y_{k}
\end{equation}
  
Where y represents the predicted result of the test data, t represents the correct result of the test data, and k represents the dimension of the data.
In addition, the ADAM is used as optimizer, since it is straightforward to implement, computationally efficient and well suited for problems with large data set and parameters.
\begin{figure} 
\centering 
\includegraphics[width=0.7\textwidth]{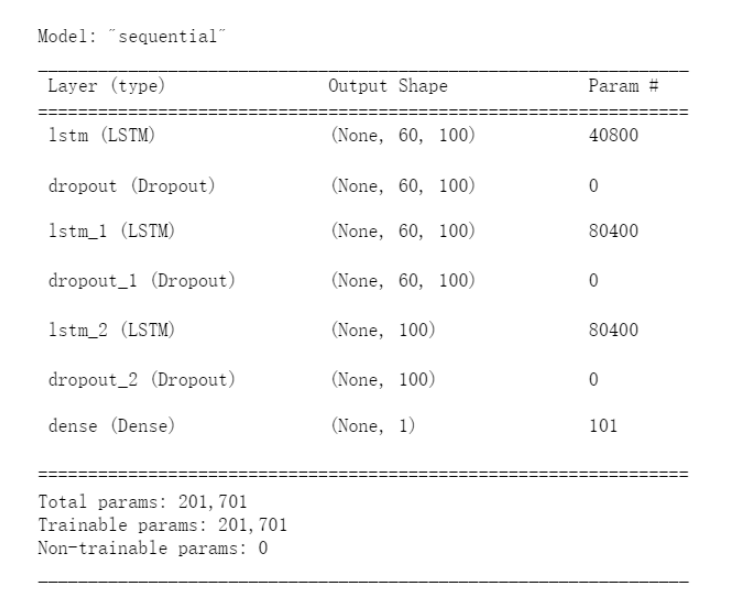} 
\caption{Model framework} 
\label{Figure.8} 
\end{figure}

\section{Result and interpretation}

The principle and construction process of ARIMA and LSTM have been introduced above. This section will show the stock price prediction results obtained by the model in this paper, and then make comparative analysis.

\subsection{Prediction of ARIMA model}
As can be seen from the above, ARIMA(1,0,1) model is used in this paper. Through the training of the model from January 2016 to December 2017, the forecast trend from January 2018 to December 2018 was obtained.

\begin{figure} 
\centering 
\includegraphics[width=0.7\textwidth]{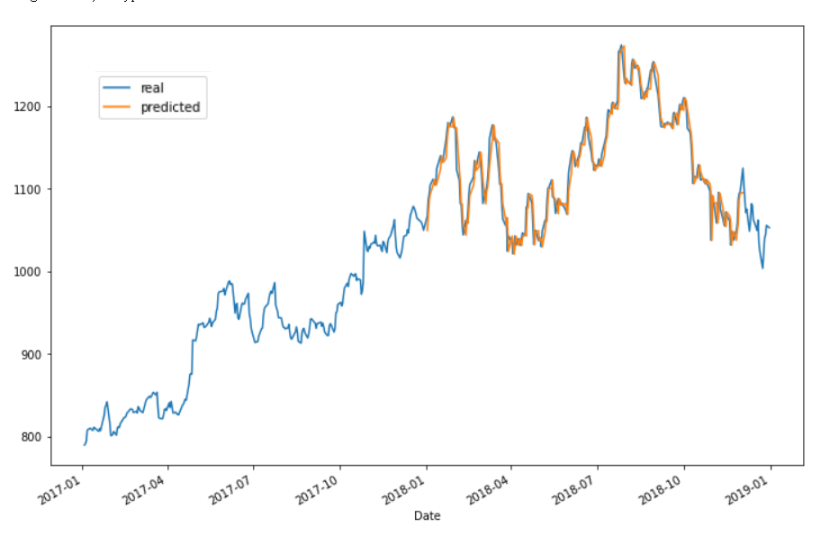} 
\caption{Prediction on Google} 
\label{Fig.9} 
\end{figure}

\begin{figure} 
\centering 
\includegraphics[width=0.7\textwidth]{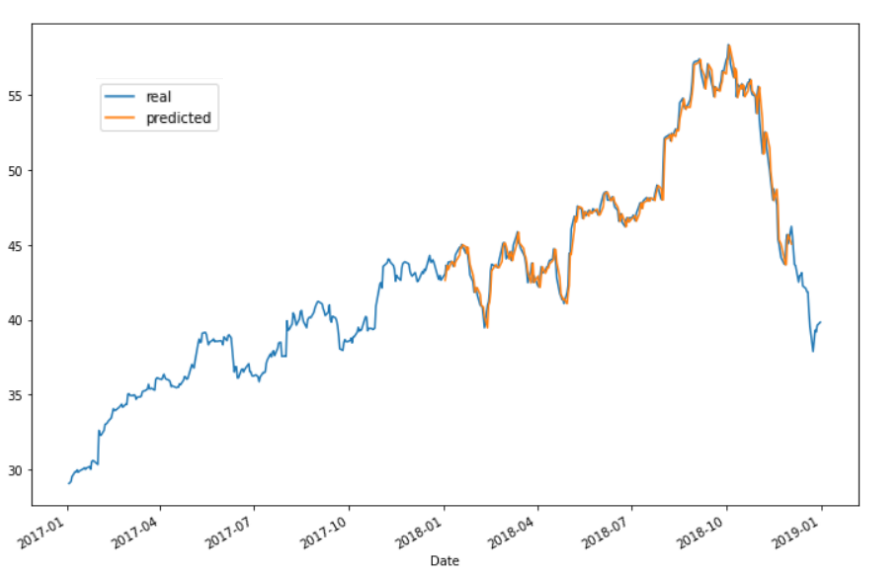} 
\caption{Prediction on Apple} 
\label{Fig.10} 
\end{figure}

\begin{figure} 
\centering 
\includegraphics[width=0.7\textwidth]{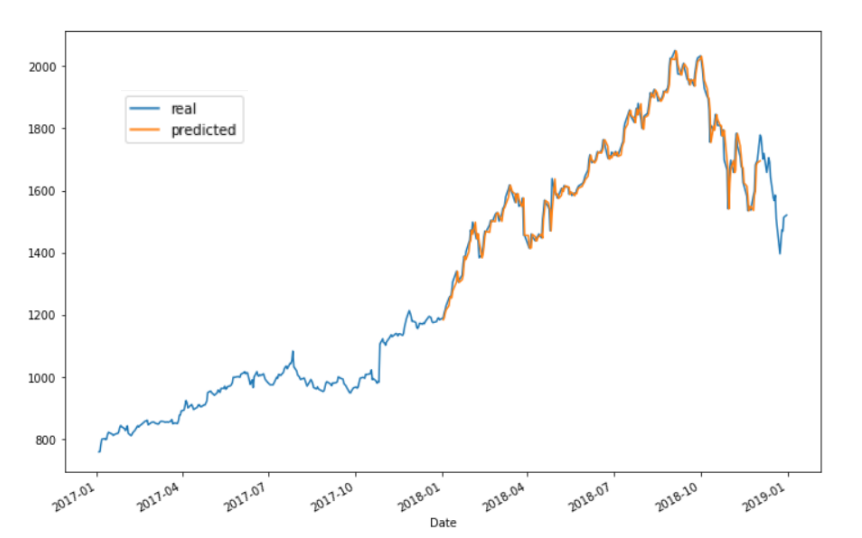} 
\caption{Prediction on Amazon} 
\label{Fig.11} 
\end{figure}

\begin{figure} 
\centering 
\includegraphics[width=0.7\textwidth]{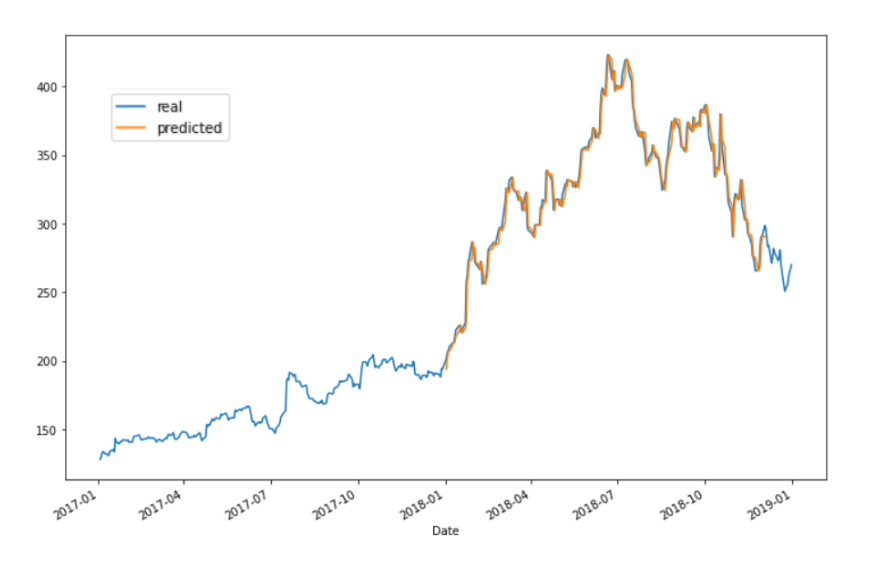} 
\caption{Prediction on Netflix} 
\label{Fig.12} 
\end{figure}

\subsection{Prediction of LSTM model}

The stock price prediction model constructed in this paper adopts multi-sequence prediction method. First, the test data is initialized into a test window with a certain sequence length of 60 days to predict the highest price of the next point. Add the actual maximum value of this point to the window, generate a new window of the same sequence length, and repeat. Among them, we set the training set from January 2010 to December 2015, the verification set from January 2016 to December 2017, and the test set from January 2018 to January 2019. It is worth mentioning that this paper tried to show the results of all data sets in one graph at a time, but different data sets were presented separately because the data set was too large to see the difference between prediction and real data. The following shows the prediction results of LSTM model constructed above for different stocks and different data sets:

\begin{figure}[htbp]
	\centering
	\subfigure[Training] {\includegraphics[width=.3\textwidth]{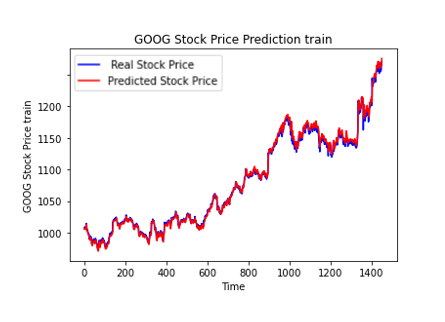}}
	\subfigure[Validation] {\includegraphics[width=.3\textwidth]{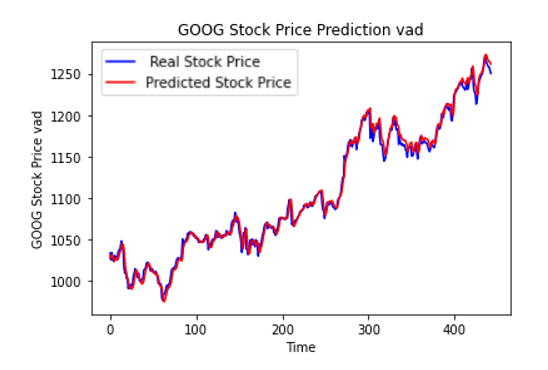}}
	\subfigure[Test] {\includegraphics[width=.3\textwidth]{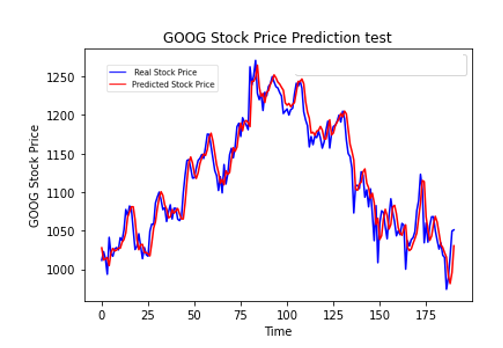}}
	\caption{Prediction on Google}
	\label{fig_E1}
\end{figure}

\begin{figure}[htbp]
	\centering
	\subfigure[Training] {\includegraphics[width=.3\textwidth]{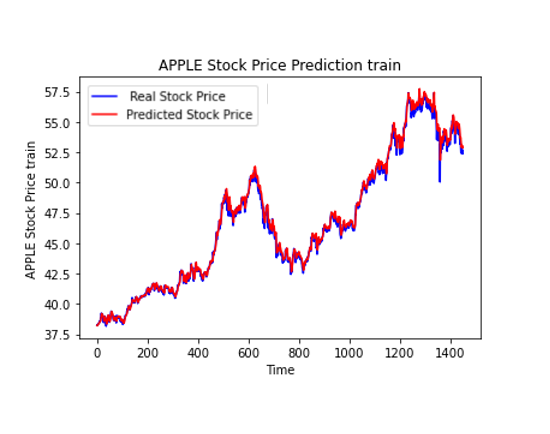}}
	\subfigure[Validation] {\includegraphics[width=.3\textwidth]{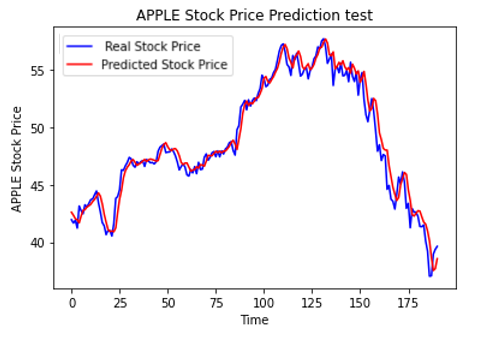}}
	\subfigure[Test] {\includegraphics[width=.3\textwidth]{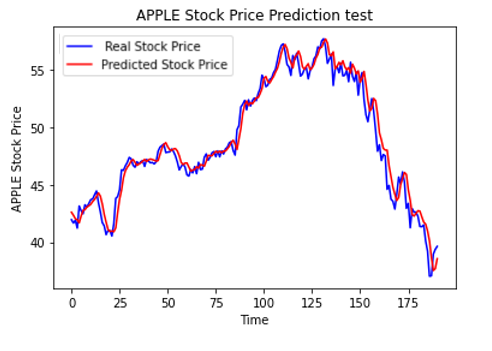}}
	\caption{Prediction on Apple}
	\label{fig_E1}
\end{figure}
\begin{figure}[htbp]
	\centering
	\subfigure[Training] {\includegraphics[width=.3\textwidth]{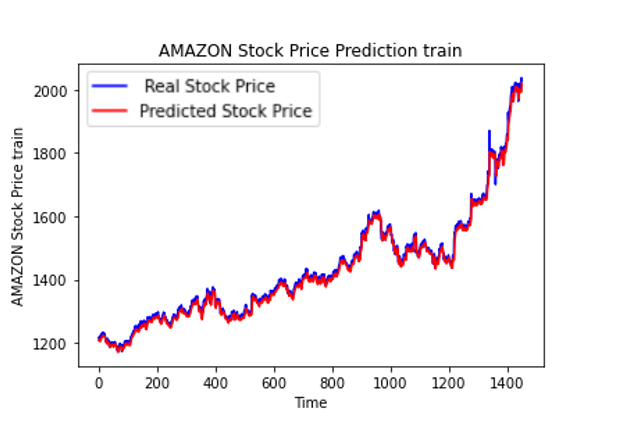}}
	\subfigure[Validation] {\includegraphics[width=.3\textwidth]{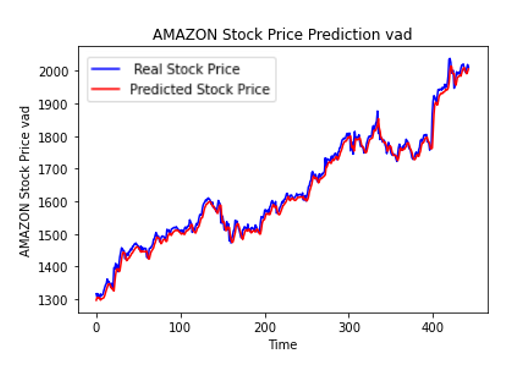}}
	\subfigure[Test] {\includegraphics[width=.3\textwidth]{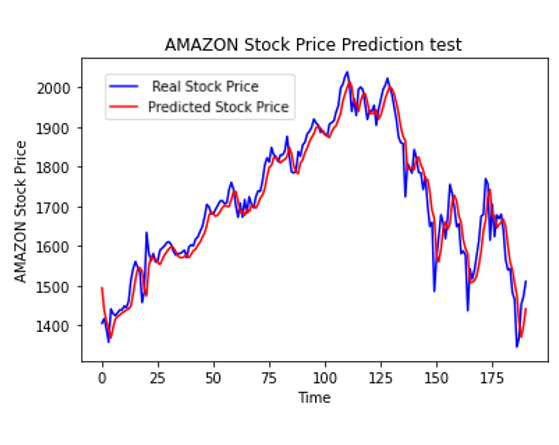}}
	\caption{Prediction on Amazon}
	\label{fig_E1}
\end{figure}
\begin{figure}[htbp]
	\centering
	\subfigure[Training] {\includegraphics[width=.3\textwidth]{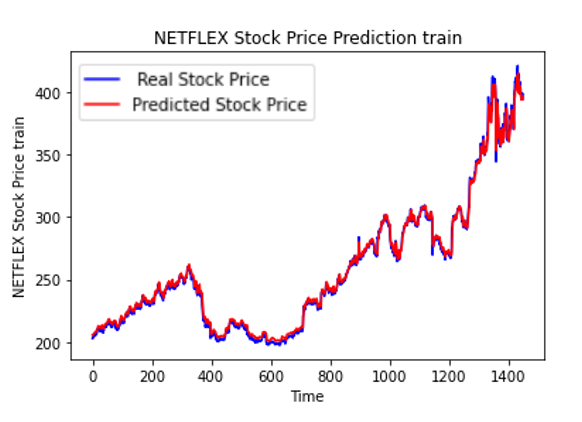}}
	\subfigure[Validation] {\includegraphics[width=.3\textwidth]{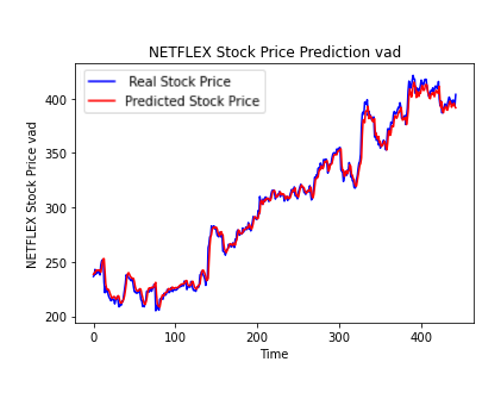}}
	\subfigure[Test] {\includegraphics[width=.3\textwidth]{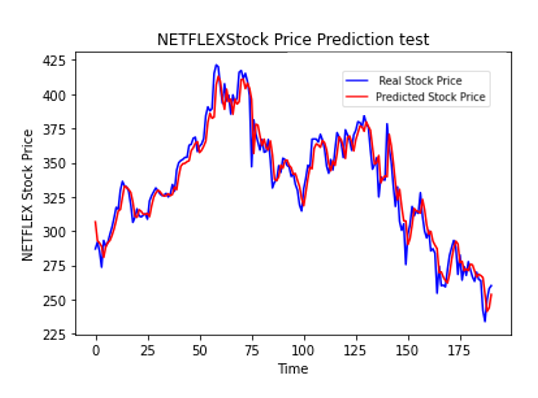}}
	\caption{Prediction on Netflix}
	\label{fig_E1}
\end{figure}

\subsubsection{Comparative analysis on different data sets}

We can see LSTM presented the excellent prediction results, among them, the visual look compared to the predictions of a training set more accurate validation set and test set, this suggests that the LSTM model has a certain degree of fitting, in the next section will introduce a new performance index, the prediction of more intuitively and accurate analysis, and the different model were analyzed.

\subsection{Comparative analysis of ARIMA and LSTM in forecast stock price}

\subsubsection{Performance evaluation metric}

In this paper, we use three methods to evaluate the performance of ARIMA model and LSTM model, which respectively are mean absolute error (MAE), mean square error (MSE) and rootmean square error (RMSE). Where f is the predicted value and y is the true value:

\begin{equation}
\begin{aligned}
\mathrm{MAE} &=\frac{1}{n} \sum_{i=1}^{n}\left|f_{i}-y_{i}\right|, \\
\mathrm{MSE} &=\frac{1}{n} \sum_{i=1}^{n}\left(f_{i}-y_{i}\right)^{2}, \\
\mathrm{RMSE} &=\sqrt{\mathrm{MSE}}
\end{aligned}
\end{equation}

\begin{table}[hbp]
 \centering
 \caption{RMSE comparison table of different models for different stocks}
 \label{tab:pagenum}
 \begin{tabular}{lllll}
  \toprule
  Stock  &Google  & Apple & Amazon &Netflix\\
  \midrule
  ARIMA &3.731571  & 4.283741 & 4.178203 &3.368191\\
  LSTM & 2.76583& 3.123431 &3.124234 &2.135342\\
  
  \bottomrule
 \end{tabular}
\end{table}

\begin{table}[hbp]
 \centering
 \caption{Evaluation of different models by different indicators}
 \label{tab:pagenum}
 \begin{tabular}{lllll}
  \toprule
      &MAE  & MSE & RMSE\\
  \midrule
  ARIMA &3.041  & 16.831 & 4.102 \\
  LSTM & 2.386& 7.272 &2.695 \\
  
  \bottomrule
 \end{tabular}
\end{table}

\subsubsection{Analysis results}
In general, although there is a certain degree of time lag between the predicted output and the real output, the simulated output of ARIMA and LSTM models for the four stock prices is basically consistent with the real output on the overall trend. In this section, MAE, MSE and RMSE performance indicators are used to analyze the prediction results of different models and different stocks.

\hspace*{\fill} \\
Firstly, LSTM is superior to LSTM in terms of the prediction performance of the two different models. This is because LSTM model can predict the future development trend of stock price only by using the evolution characteristics of the historical state of stock price itself. Compared with traditional mathematical statistics method, LSTM has the characteristics of simplicity and reliability. The main reason is that the stock price historical change state is itself characteristic and the result of comprehensive function of many factors, each stock price changes in a certain time point, and implied in the time series of the whole historical process, a moment before the foundation of growth and decline of the stock price point of the next moment, both maintained close relations, to maintain the continuity of the whole stock price evolution, LSTM just reveals the internal law of stock price changes, so it is feasible and effective to predict stock price changes by LSTM. Stock price changes and the political, economic, investor psychology, closely related to the enterprise management status, and so on a variety of factors, especially by is extremely remarkable economic effect, cause the price difference between different time has great randomness and chance, sometimes even haphazard, it brings to the stock price prediction of great hardship, But LSTM model can give the change of stock price, especially for the short-term prediction of stock price has a certain reference significance. However, ARIMA time series method is not as good as LSTM in mastering these memorized data.

\hspace*{\fill} \\

Secondly, from the difference of different stocks, the prediction effect of Google and Netflix is better than that of the four simulation prediction experiments, while the deviation between the predicted value and the real value of Apple and Amazon is larger, and the prediction effect is slightly inferior. This shows that the prediction effect of the model is different for different stocks. The prediction effect of individual stocks is not only related to stock market factors, but also related to the factors of individual stocks themselves.

\hspace*{\fill} \\

Finally, from the perspective of the training complexity and time of ARIMA and LSTM, although the performance of ARIMA is inferior to LSTM, its training time is short and training parameters are few, making it easier to draw conclusions.

\hspace*{\fill} \\
\section{Ideas for Future Work}
In this paper, MAE, MSE and RMSE performance indicators are used to analyze the performance of different stocks predicted by LSTM and ARIMA models, and the following conclusions can be drawn:
(1)	Both ARIMA and LSTM models can predict stock prices, and the prediction results are generally consistent with the actual results.
(2)	Compared with ARIMA, LSTM has better performance in predicting stock prices, while ARIMA is inferior to LSTM in expressing stock price changes. 
(3)	Both ARIMA and LSTM have different prediction effects on different stocks even if the same model is used. 
(4)	Although ARIMA's stock price prediction performance is not as good as LSTM's, its training time is short, training parameters are few, and its application is more convenient.

Additionally, the shortcomings of the article are that, first, the article only tests and analyzes the ARIMA and LSTM models, and finds that there is a certain time lag between the predicted results of the stock price and the actual results of both models; Second, because the dataset is too large to show the results of all the datasets in the same graph, the difference between the predicted and real data is not intuitive enough. Therefore, on this basis, future research can consider making efforts from two aspects, testing, and analyzing models other than ARIMA and LSTM, and improving or solving the problem of stock price forecasting time lag; And consider other techniques or methods to make the differences between predicted and real data more intuitive.

\bibliographystyle{unsrt}  
\bibliography{references}

\end{document}